\documentclass[11pt]{article}

\usepackage[a4paper,margin=2.5cm]{geometry}

\usepackage{graphicx}
\usepackage{subcaption}
\usepackage[percent]{overpic}

\usepackage{amsmath,amssymb}
\usepackage{natbib}

\usepackage{hyperref}
\usepackage{microtype}   
\usepackage{setspace}    
\usepackage{titlesec}    
\usepackage{authblk}     

\newcommand\BibTeX{{\rmfamily B\kern-.05em \textsc{i\kern-.025em b}\kern-.08em
T\kern-.1667em\lower.7ex\hbox{E}\kern-.125emX}}

\usepackage{moreverb}
\usepackage{gensymb}
\usepackage{graphicx}

\usepackage{subcaption}
\usepackage[percent]{overpic}

\onecolumn
\usepackage[font=small,labelfont=bf]{caption}
\begin{document}

\title{Hybrid ensemble forecasting combining physics-based and machine-learning predictions through spectral nudging
}

\author{
\textbf{Inna Polichtchouk}$^{1}$\thanks{Corresponding author: inna.polichtchouk@ecmwf.int},
\textbf{Simon Lang}$^{1}$,
\textbf{Sarah-Jane Lock}$^{1}$,
\textbf{Michael Maier-Gerber}$^{1}$,
\textbf{Peter Dueben}$^{1}$
}

\date{$^{1}$European Centre for Medium-Range Weather Forecasts (ECMWF), Reading, UK}

\maketitle
\begin{abstract}
 We present the first application of spectral nudging in a probabilistic ensemble forecasting framework, combining the physics-based ECMWF Integrated Forecasting System ensemble (IFS-ENS) with forecasts from the probabilistic machine-learned AIFS-ENS ensemble. Large scales of virtual temperature and vorticity are relaxed toward the machine-learned forecasts, while mesoscale structures remain governed by the physics-based model. This hybrid ensemble shows substantial improvements in large-scale forecast skill, with gains in predictive skill extended by up to two days in the tropics and by approximately half a day in the extra-tropics relative to IFS-ENS. Despite nudging being applied only to upper-air fields, improvements are also found in several near-surface parameters. Tropical cyclone track forecasts improve significantly, consistent with improved representation of the large-scale steering flow, without degrading storm intensity or ensemble spread. These results demonstrate that spectral nudging can be successfully extended to ensemble prediction and provide an attractive pathway for combining machine-learned and physics-based weather prediction systems.
\end{abstract}

\section{Introduction} 

Machine-learned deterministic and probabilistic global numerical weather prediction (NWP) models developed by research labs and technology companies are now outperforming traditional physics-based models in large-scale forecast skill, as measured by metrics such as root-mean-square error, anomaly correlation, mean bias, and continuous ranked probability score \citep[e.g.,][]{benboullage2024,chen2023fuxi,lam2023learning,lang2024aifsA,price2025probabilistic,LangAIFSENS2024}. The first generation of machine-learned models are typically trained deterministically to minimize mean squared error, which encourages smoothing of forecast fields over time and a reduction in forecast activity when applied in an auto-regressive manner \citep{PhilippeMaierGerber}. The smoothing helps to reduce the so-called “double penalty” that arises when misplaced fine-scale features are penalized twice \citep[e.g.,][]{erbert2013,LangAIFSENS2024}. However, this smoothing tendency often results in overly diffuse small-scale structures, limiting the models’ ability to accurately capture features such as tropical cyclone intensities \citep[e.g.,][]{liu2024evaluation,demaria2024evaluation} and orographic precipitation extremes \citep{pan2025comparative}. State-of-the-art probabilistic machine-learned models are not prone to such smoothing \citep[e.g.,][]{LangAIFSENS2024,price2025probabilistic} but are currently constrained by their relatively coarse resolution, which hinders the representation of high-impact features compared to operational physics-based NWP systems.

To combine the superior large-scale skill of machine-learned models with the more physically consistent small-scale realism of physics-based NWP models,  \cite{husain2024leveraging} proposed a hybrid approach for deterministic forecasting at Environment and Climate Change Canada (ECCC). In their setup, the large scales of the GEM (Global Environmental Multiscale) model were constrained using forecasts from the machine-learned GraphCast model via a spectral nudging technique. Specifically, GEM’s virtual temperature and horizontal winds in the 850–250 hPa layer were nudged toward GraphCast forecasts, but only for horizontal wavelengths larger than 2000 km. This hybrid configuration significantly improved large-scale skill relative to GEM forecasts — yielding gains in predictive skill of up to 34 hours — and positively impacted precipitation forecasts and tropical cyclone track errors, while leaving storm intensities largely unchanged. A similar approach has since been successfully tested at the European Centre for Medium-Range Weather Forecasts (ECMWF) within the Integrated Forecasting System (IFS), using a variant of the Artificial Intelligence Forecasting System (AIFS) that predicts on model levels \citep{Polichtchouk2025}.

While these results demonstrate the potential of spectral nudging in deterministic forecasting, extending this approach to ensemble prediction systems introduces additional challenges. In a probabilistic framework, spectral nudging interacts directly with ensemble spread, flow-dependent uncertainty, and probabilistic reliability, and therefore cannot be viewed as a trivial extension of the deterministic case. In particular, constraining the large-scale flow aligns the ensemble spread of the hybrid model with the ensemble spread of the machine-learned model, and the key requirement is that such changes remain consistent with corresponding changes in forecast error.

To date, the application of spectral nudging in a probabilistic hybrid forecasting framework has not yet been explored. Here, we present the first application and systematic evaluation of spectral nudging in a global ensemble prediction system. We extend the spectral nudging methodology to the ECMWF IFS ensemble (IFS-ENS), using the probabilistic AIFS-ENS model \citep{LangAIFSENS2024} as the machine-learned reference. By nudging each IFS ensemble member toward a corresponding probabilistic machine-learned forecast member, we investigate whether improvements in large-scale forecast skill can be achieved without degrading ensemble forecast skill, and whether the benefits previously demonstrated for deterministic hybrid systems extend to probabilistic forecasts, including extreme events.

In this study, the machine-learned ensemble provides flow-dependent large-scale information that constrains the evolution of the physics-based ensemble, while scales below the nudging cut-off continue to evolve under the physically based dynamics and parametrizations of the NWP system. From this perspective, hybrid forecasting provides a way to incorporate machine-learned large-scale skill within an existing high-resolution operational framework. 
This scale separation allows machine-learned skill to be used where it is strongest, while retaining decades of development and verification of resolved-scale behaviour in physics-based models.

The remainder of this paper is structured as follows. Section~\ref{sec:method} describes the probabilistic IFS-ENS and AIFS-ENS models, along with the spectral nudging methodology. In Section~\ref{sec:results}, we assess the forecast skill of the hybrid IFS-ENS relative to the physics-based IFS-ENS, using 15-day forecasts at 9 km resolution. Here, a comparison to the pure machine-learned AIFS-ENS ensemble is also presented. This section also examines tropical cyclone track and intensity predictions in the hybrid system as well as the overall impact of hybrid forecasting on the extreme near-surface events in the Northern Hemisphere. Finally, Section~\ref{sec:conclusion} presents a summary, discussion and conclusions.

\section{Models and Methods}\label{sec:method}
\subsection{ECMWF IFS}
We use the ECMWF Integrated Forecasting System (IFS) ensemble prediction system \citep{molteni1996ecmwf, leutbecher2008ensemble}, in the operational CY49R1 configuration \citep{Roberts_49r1}, run at TCo1279 horizontal resolution \citep{lang81380}, corresponding to a spectral truncation at total wavenumber 1279 and an average grid-spacing of 9~km. The IFS is coupled to the 1/4$\degree$ NEMO ocean model and employs a semi-implicit, semi-Lagrangian formulation \citep{hortal2002development,temperton2001two,diamantakis2022fast} and is horizontally discretized using a spherical harmonic spectral expansion, combined with a cubic-octahedral reduced Gaussian grid \citep{Malardel2016}. Time integration is performed with a time step of $\Delta t$=450s. Vertically, the IFS uses a pressure-based hybrid $\eta$-coordinate with 137 model levels extending from the surface up to 0.01 hPa. The vertical discretization uses a third-order finite element method. 

We perform 50-member ensemble forecasts out to 15 days. Ensemble initial condition perturbations are generated using a combination of singular vectors and the Ensemble of Data Assimilations (EDA), while model uncertainty is represented using the Stochastically Perturbed Parameterizations scheme (SPP) \citep{ollinaho2017towards, leutbecher2017, langspp, leutbecher2024improving}, applied throughout the forecast integration. To ensure a consistent comparison, all three forecast systems — the standard IFS ensemble (IFS-ENS), the spectrally nudged hybrid IFS ensemble (hy-IFS-ENS), and the AIFS ensemble used for nudging (see section~\ref{sec:aifs-crps-ml}) — are initialized from identical initial conditions. The initial perturbations are generated using a combination of singular vectors and the Ensemble of Data Assimilations (EDA), and are applied equally across all systems (see \cite{lang2021more} for a description of the initial perturbations configuration in the IFS). This design ensures consistent initial conditions across ensemble members of the machine-learned and hybrid systems, thus reducing competing perturbation structures between pairs of IFS and AIFS members and mitigating potential artefacts arising from nudging one towards the other. In addition, differences in forecast performance can be attributed solely to the modeling approach — whether physics-based, machine-learned, or hybrid — rather than to differences in initial conditions.

\subsection{Probabilistic Model-Level AIFS-ENS}\label{sec:aifs-crps-ml}
The AIFS-ENS model is a machine-learned, probabilistic global forecasting system. It is trained with an objective that is based on the Continuous Ranked Probability Score (CRPS, see for example \cite{hersbach2000decomposition} and \cite{Ferro2014}). This tasks the model to produce skillful ensemble forecasts, in which each ensemble member maintains a realistic level of variability. AIFS-ENS is built on an encoder–processor–decoder architecture. The encoder maps the high-resolution input state to a lower-resolution latent space using a Graph Neural Network (GNN). A sliding window transformer-based processor then forecasts the evolution of this latent state 6 hours forward in time. The decoder projects the updated latent state back to the target resolution, which corresponds to a reduced Gaussian grid at spectral resolution O96 (approximately 120 km), in contrast to the operational AIFS-ENS configuration, which runs at N320 resolution (approximately 28 km). Further architectural details are described in \cite{lang2024aifsA} and \cite{LangAIFSENS2024}.

While the operational AIFS-ENS predicts atmospheric fields on 13 pressure levels, we employ an experimental model-level version trained to predict variables, such as wind and temperature, on IFS model levels 137 (at the surface) to 50 (at approximately 56~hPa). Hereafter, this configuration is denoted AIFS-ENS-ML to distinguish it from the operational AIFS-ENS. This enables seamless integration with the IFS, eliminating the need for vertical interpolation at each step and thereby improving computational efficiency and consistency. The AIFS-ENS-ML was initially trained on ERA5 reanalysis data spanning 1979–2017, and later fine-tuned using ECMWF’s operational high-resolution analyses, for the period 2016–2023. For training AIFS-ENS-ML, all data is up-scaled to an O96 grid. 

\subsection{Spectral nudging}
The spectral formulation of the IFS makes it straightforward to adopt spectral nudging. In this study, we apply nudging to the prognostic spectral coefficients of virtual temperature ($T_v$) and vorticity ($\zeta$), constraining only the large scales corresponding to total wavenumbers 0–20 (approximately corresponding to horizontal wavelengths larger than 2000~km at the equator; T21 truncation).

Nudging is applied by adding an extra relaxation tendency on the right-hand side of the prognostic equations for $T_v$  and  $\zeta$, as follows:
\begin{equation}
    \frac{\partial X}{\partial t}=... -\frac{f(t) g(k)}{\tau}(X_{\text{ifs}}-X_{\text{aifs}}),
\end{equation}
where $X \in \{T_v,\zeta\}$; $\tau$ is the relaxation time-scale (set to 12 hours, following \cite{husain2024leveraging} and confirmed here as optimal), $f(t)$ is a temporal ramp function, and $g(k)$ is a vertical mask that restricts nudging below the tropopause. The restriction to the troposphere is motivated by the fact that current medium-range machine-learned models do not exhibit systematically improved skill in the stratosphere relative to the physics-based IFS. Nudging the large-scale flow to AIFS-ENS above the tropopause could therefore degrade stratospheric forecast skill.  This formulation allows the large-scale flow to be constrained by the machine-learned ensemble system while preserving the internally generated smaller scale variability of the IFS.

The vertical mask $g(k)$ is defined as:
\begin{equation}
    g(k)=\frac{1}{1+\exp(k_{\text{trp}}+5-k)},
\end{equation}
where $k = 1,\dots,137$ is the model level index, and $k_{\text{trp}}$ denotes the level of the lapse-rate tropopause, which varies in space and time. The temporal ramp function $f(t)$ is given by:
\begin{equation}
    f(t)=\frac{1}{2}\left( \tanh\left[\frac{t} {t_{\text{ramp}}} - 1 \right ] + 1.0 \right),
\end{equation}
where $t$ is continuous forecast time, with $t=j \Delta t$ in the discrete formulation; $\Delta t = 450 \, \text{s}$ is the IFS time step size, and $t_{\text{ramp}}=8640$~s.  This time ramp gradually increases the nudging strength over the first 8–12 hours of the forecast, reaching full strength at around 12 hours. This approach mitigates degradation in early forecast skill against analysis, as the free-running IFS typically performs better than the nudged configuration immediately after initialization.

The nudging is applied at every model time step. Since AIFS-ENS-ML predictions are only available at 6-hourly intervals, linear temporal interpolation is used to compute  $X_\text{aifs}$  at intermediate times. In practice, a 15-day AIFS-ENS-ML forecast is first run, with output saved every 6 hours for specific humidity ($q$), temperature ($T$), and horizontal wind components $(u, v)$ on model levels 50-137. The virtual temperature is then computed in grid-point space as  $T_v = T (1 + 0.61q)$. This  $T_v$  field, along with  $(u,v)$, is transformed into spectral space to obtain the spectral coefficients of virtual temperature and vorticity, which are used as the nudging target fields.  This approach ensures that the nudging acts consistently on prognostic variables within the native IFS spectral framework.

We found that nudging the zonal wavenumber  $m = 2$  of divergence introduced aliasing of the semi-diurnal tidal signal. This resulted in a wavenumber-2 pattern of degradation in tropical forecasts of mean sea level pressure and geopotential height. To avoid this artefact, divergence is excluded entirely from the nudging procedure. While this choice limits direct control over divergent large-scale modes, it avoids spurious dynamical responses associated with tidal aliasing.

Once the nudging input fields are prepared, an IFS-ENS-ML forecast is run with spectral nudging activated. Each perturbed member of IFS is nudged to the corresponding member from AIFS-ENS-ML, both of which are initialized from identical initial conditions. This member-to-member nudging preserves the ensemble structure of the large-scale flow and avoids an artificial collapse of ensemble spread.

One might try to generate an ensemble from deterministically trained machine-learned forecast models (e.g., AIFS-Single) using perturbed initial conditions. Here, one would make the assumption that the resulting ensemble system provides sufficient large-scale spread for the nudging, with the remaining uncertainty developing through SPP at smaller scales. In practice, however, such an approach does not lead to adequate spread. A lack of spread, even at large scales, is a well-known property of machine-learned models trained with regression losses such as MSE \citep[e.g.][]{mahesh2025huge}. The nudging constrains the large scale evolution of the physics-based model to the large scale evolution of the machine-learned forecast model. AIFS-ENS-ML is probabilistically trained and hence provides a more consistent and explicitly probabilistic representation of large-scale uncertainty. Member-to-member nudging transfers this large-scale ensemble uncertainty to hy-IFS-ENS, while SPP allows additional uncertainty to develop at smaller scales. This combination makes it possible to build a skilful and reliable hybrid system.

\subsection{Experimental setup}
\begin{figure}
\centering
\includegraphics[width=\textwidth]{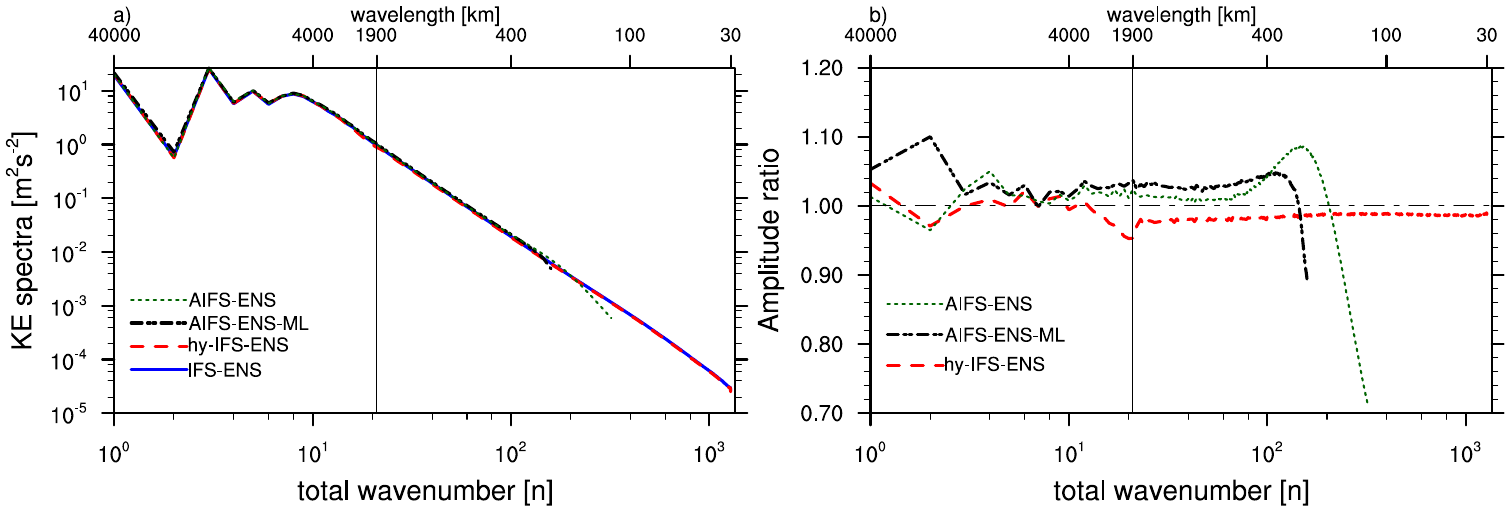}
\caption{(a) Global kinetic energy spectra as a function of total wavenumber at 500~hPa for forecast day 10 of the model-level AIFS-ENS-ML (dash-dotted black), IFS-ENS (solid blue), hy-IFS-ENS (dashed red) and operational AIFS-ENS (dotted green). (b) Spectral amplitude ratios relative to IFS-ENS, defined as the square root of the ratio of the corresponding kinetic energy spectra \citep{husain2024leveraging}.  Spectra are averaged for all 50 perturbed members and for all forecasts initialized at 00 UTC between 1 January and 28 February 2026. Black vertical line shows total wavenumber 21, which is the cut-off wavenumber for nudging in hy-IFS-ENS. AIFS-ENS-ML is shown on the model level closest to 500~hPa, while the other systems are shown at 500~hPa.}\label{fig:fig1}
\end{figure}

Nudged and un-nudged 15-day ensemble forecasts, each with 50 perturbed members, are initialized every day at 00UTC from 20 October 2025 to 12 May 2026. Standard probabilistic forecast skill metrics are computed using the in-house verification software QUAVER. Scores against ECMWF analysis are computed on a 1.5$\degree \times$ 1.5$\degree$ grid. For verification against radiosonde and SYNOP observations, full-resolution model fields are interpolated to observation locations using nearest-neighbour interpolation (O1280 for IFS-ENS and hy-IFS-ENS; N320 for operational AIFS-ENS). Tropical cyclones are tracked using the method described in \cite{vitart2012new} and \cite{magnusson2021tropical}.

To demonstrate that spectral nudging does not distort the spatial variability of the forecast fields, Figure~\ref{fig:fig1}a shows kinetic energy spectra at 500~hPa at 10-day lead time for the IFS-ENS, spectrally nudged IFS-ENS (hy-IFS-ENS), operational AIFS-ENS and the model-level, low-resolution AIFS-ENS-ML used for nudging. The spectra are nearly indistinguishable across all systems at resolved scales, including mesoscale wavelengths (wavenumbers up to $\sim$200, corresponding to horizontal wavelengths of approximately 200 km). However, the spectral amplitude ratios relative to IFS-ENS (Figure~\ref{fig:fig1}b), defined following \cite{husain2024leveraging}, reveal small differences. In particular, hy-IFS-ENS has slightly lower spectral amplitude than IFS-ENS, by approximately 1-2\% over much of the spectrum, including at wavenumbers beyond those directly nudged.

Within the directly nudged scales, the spectral amplitude of hy-IFS-ENS does not exactly follow that of the AIFS-ENS-ML nudging target. This may partly reflect the 12-h relaxation timescale, which allows the IFS dynamics to contribute to the evolution of the nudged scales rather than tightly constraining them to the AIFS-ENS-ML target. In addition, AIFS-ENS-ML output is available at 6-hourly intervals and is linearly interpolated to each IFS time step, which may introduce some temporal smoothing. The AIFS-ENS-ML spectrum is also evaluated on the model level closest to 500~hPa rather than exactly at 500~hPa, and therefore does not provide an exact like-for-like comparison. Overall, the close agreement of the spectra indicates that spectral nudging largely preserves the spatial variability of IFS-ENS.

In deterministic hybrid systems, nudging beyond wavenumber 21 risks introducing excessive smoothing, since deterministic machine-learned models tend to suppress mesoscale variability. Probabilistic models such as AIFS-ENS do not exhibit this behaviour (see Figure~\ref{fig:fig1}) and could, in principle, support nudging at higher wavenumbers. We tested cut-off wavenumbers T42 and T85 in addition to T21 in coarser-resolution TCo399 nudged ensemble experiments. Nudging to T42 yielded only marginal further improvements (typically 1–2\% for upper-air variables), while T85 provided no additional benefit. It remains to be tested whether training AIFS-ENS-ML at higher N320 resolution would make nudging to T42 or T85 more beneficial, and whether, in full-resolution TCo1279 experiments, higher cut-off wavenumbers could provide additional benefit for more localized phenomena such as tropical-cyclone track forecasts. We therefore adopt T21 for this study as a conservative and robust choice that limits the degree of machine-learned intervention on the physics-based model. 

\begin{figure}
\centering
\includegraphics[width=0.85\textwidth]{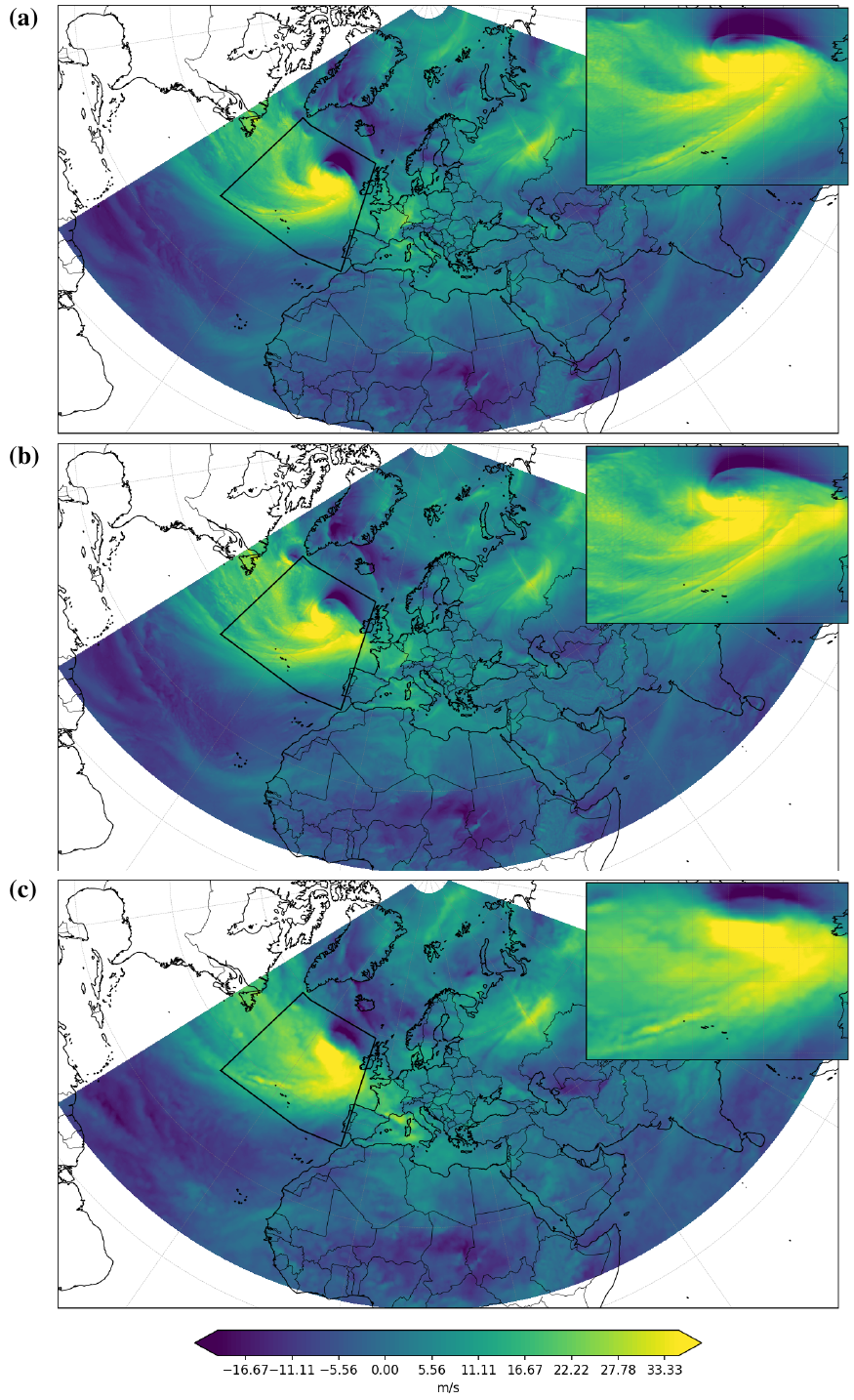}
\caption{850\,hPa meridional wind ($v$ component; m\,s$^{-1}$) at forecast day 4 for ensemble member one of (a) IFS-ENS, (b) hy-IFS-ENS, and (c) operational AIFS-ENS initialized on 20 January 2025 at 00 UTC.}
\label{fig:fig_inlet}
\end{figure}

\section{Results}\label{sec:results}
\subsection{Example of hybrid forecast structure}
Figure~\ref{fig:fig_inlet} shows an illustrative example of the 850 hPa meridional wind at forecast day 4 for IFS-ENS, hy-IFS-ENS, and AIFS-ENS during wind storm Eowyn. The example corresponds to a single ensemble member and is shown for visual comparison only. The hy-IFS-ENS exhibits spatial structure and levels of detail comparable to the physics-based IFS-ENS, with no visible reduction in spatial variability despite the application of spectral nudging. This illustrates that the nudging primarily affects the large scales of the flow while leaving the resolved smaller scale spatial structure of the ensemble forecast unchanged.

\subsection{Large-scale skill scores}\label{sec:large-scale}
\begin{figure}
\centering
\includegraphics[width=\textwidth]{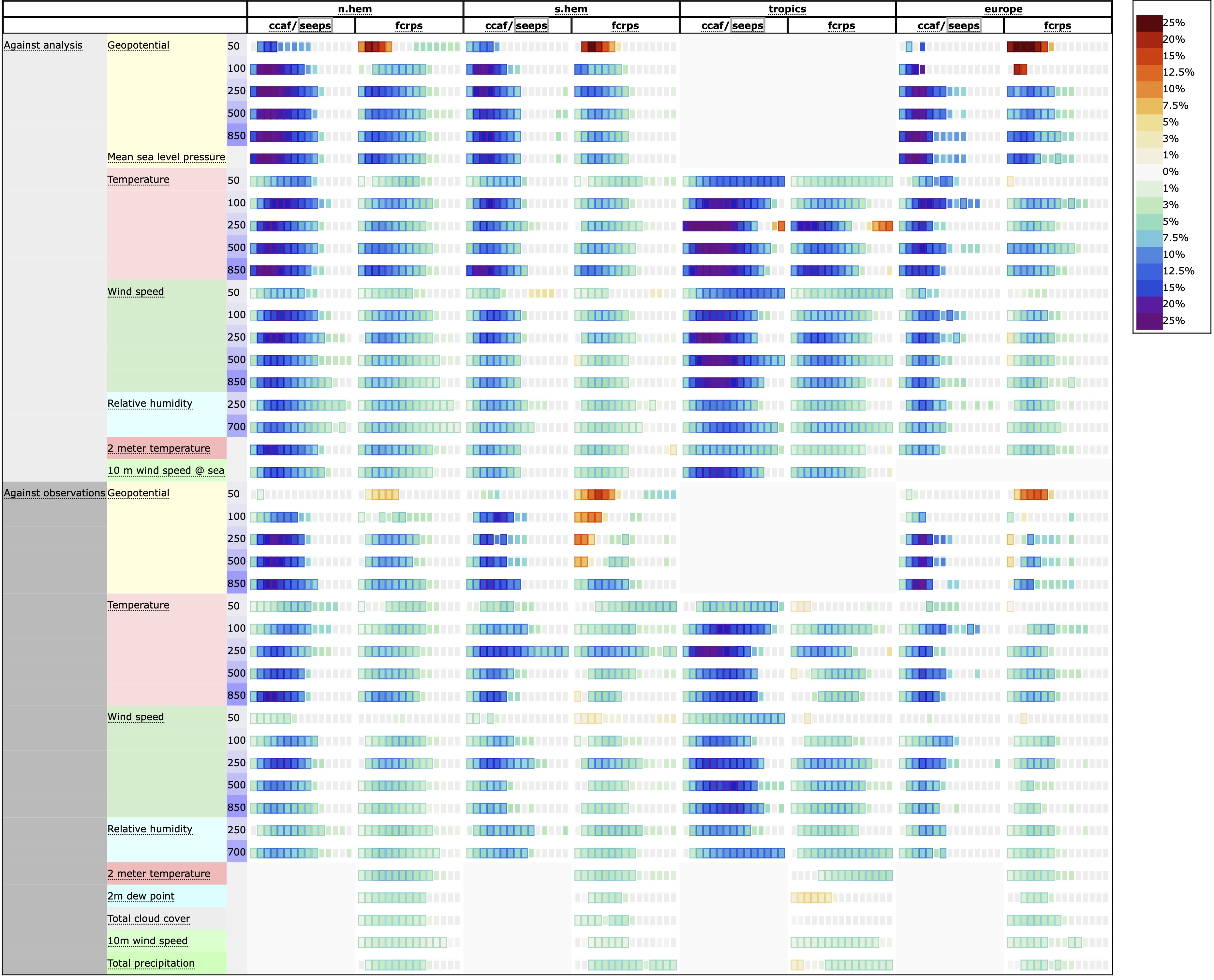}
\caption{Summary scorecard comparing the difference in forecast skill between the spectrally nudged IFS ensemble (hy-IFS-ENS) and the control IFS ensemble (IFS-ENS), using the fair continuous ranked probability score (FCRPS) and the anomaly correlation of the ensemble mean (CCAF), verified against ECMWF operational analysis (upper half) and against radiosonde and SYNOP station observations (lower half). Each box corresponds to a 24-h lead-time interval from forecast day 1 to day 15. Shades of blue indicate improvement in hy-IFS-ENS relative to IFS, and shades of red indicate degradation. Colour saturation reflects the magnitude of the normalized difference. Statistically significant differences at the 99.7\% confidence level are outlined with coloured frames. Regional abbreviations denote Northern Hemisphere (n.hem; 20$\degree$N-90$\degree$N), Southern Hemisphere (s.hem; 20$\degree$S–90$\degree$S), Tropics (20$\degree$N-20$\degree$S) and Europe (35$\degree$N-75$\degree$N, 12.5$\degree$W-42.5$\degree$E). Scores are computed using 50 perturbed members from 207 forecasts initialized at 00 UTC between 20 October 2025 and 12 May 2026.}
\label{fig:fig2}
\end{figure}

Figure~\ref{fig:fig2} presents a summary scorecard for selected variables averaged over several regions, comparing the spectrally nudged hy-IFS-ENS against the standard IFS-ENS. Verification is shown against the ECMWF operational analysis (upper half) and against radiosonde and SYNOP surface station observations (lower half). Forecast skill is evaluated using the fair continuous ranked probability score (FCRPS). FCRPS accounts for both, resolution and reliability, and provides an unbiased estimate of probabilistic skill that is appropriate for small ensemble sizes \citep{Ferro2014}.
In addition, the anomaly correlation of the ensemble mean (CCAF) is shown for completeness.

Overall, nudging the IFS-ENS toward AIFS-ENS yields a predominantly blue scorecard, indicating widespread improvements in forecast skill. Blue shading denotes that hy-IFS-ENS outperforms IFS-ENS. The most substantial gains occur in the tropics, where upper-air scores for temperature and wind improve by up to 25\%, particularly in the 850–250 hPa layer. This is notable, as it corresponds to forecast skill improvements that would take many years of physics-based model developments. Hence, the results highlight the potential of this hybrid approach to combine machine-learned and physics-based models for improved forecast skill.  Additional experiments run with a smaller eight-member ensemble for the full summer and winter season show that nudging provides larger benefits in the summer hemisphere compared to the winter hemisphere (not shown). Improvements in surface parameters — including 2-m temperature and 10-m wind speed over the ocean — reach up to 15\% when verified against the ECMWF operational analysis. However, when verified against SYNOP surface station observations, the improvements are smaller, typically in the range of 1–3\%. 

Interestingly, although the spectral nudging is applied only below the tropopause, hy-IFS-ENS also shows improved forecast skill in the stratosphere. This is evident across most stratospheric fields at both 100~hPa and 50~hPa, with the exception of the FCRPS for geopotential height at 50~hPa, which shows some degradation due to a small shift in the mean bias. These improvements are consistent with a more accurate troposphere providing improved lower-boundary forcing for the stratosphere.

In addition to the degradation in FCRPS for geopotential height in the stratosphere, the 2-metre dewpoint temperature verified against SYNOP observations in the tropics is degraded by about 2\% during the first 6 days of the forecast, due to an increase in the mean absolute bias. In the tropics, at forecast lead times beyond 12 days, there is also some degradation in mid-tropospheric temperature verified against ECMWF analysis. This appears to be seasonal: it is present in the winter months but absent in the spring months (not shown).

\begin{figure}
\centering
\includegraphics[width=\textwidth]{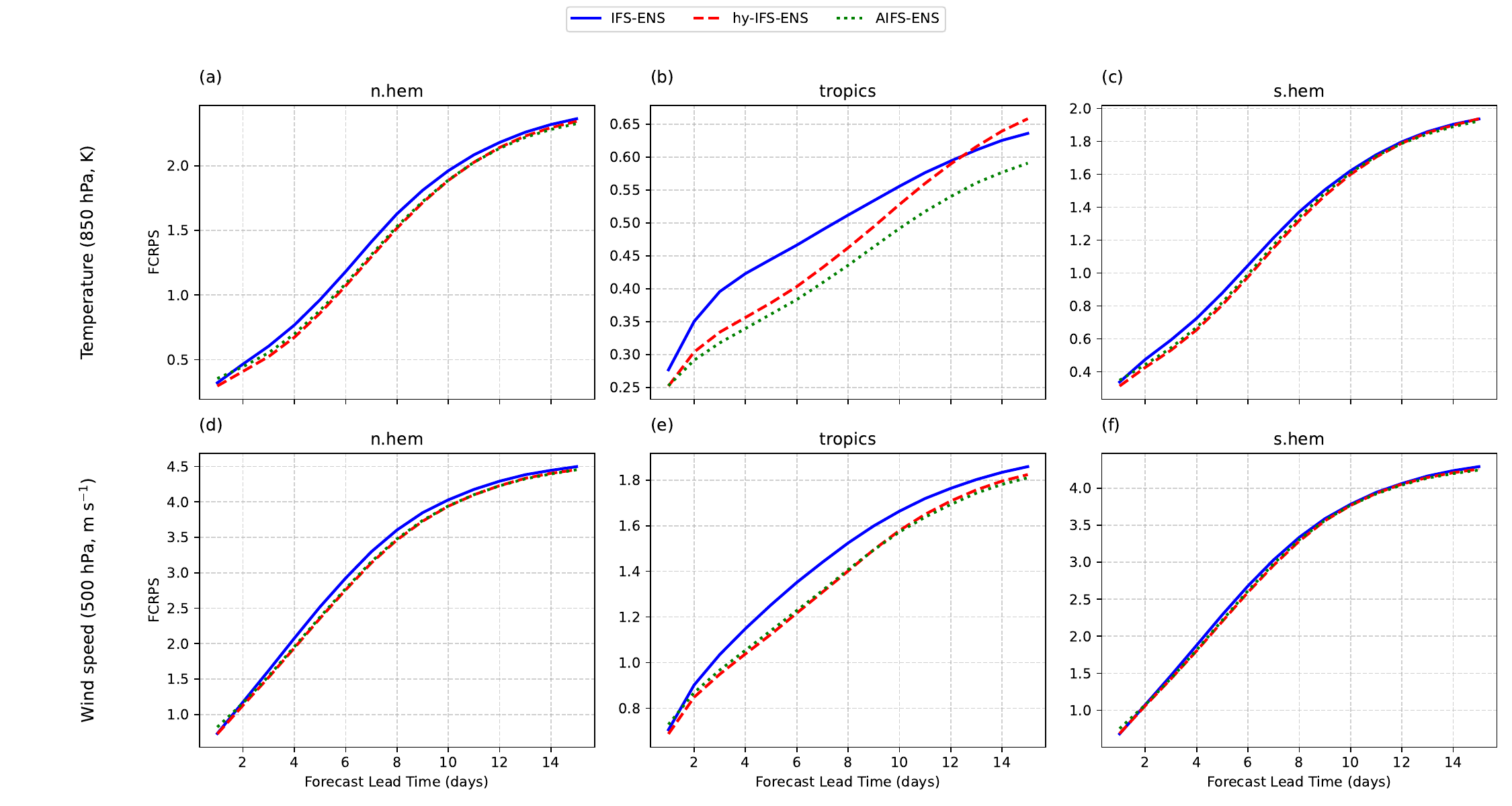}
\caption{Fair CRPS for (a,b,c) 850 hPa temperature and (d,e,f) 500 hPa wind speed for hy-IFS-ENS (dashed red), IFS-ENS (solid blue) and operational AIFS-ENS (dotted green) over the (a,d) Northern Hemisphere, (b,e) tropics and (c,f) Southern Hemisphere. For differences between IFS-ENS and hy-IFS-ENS, the data here correspond to the lines labelled ``Temperature 850" and ``Wind speed 500" in Figure~\ref{fig:fig2}.  Smaller FCRPS values indicate higher skill. Scores are computed against ECMWF operational analysis from 207 forecasts initialized at 00 UTC between 20 October 2025 and 12 May 2026. }
\label{fig:fig3}
\end{figure}

\begin{figure}
\centering
\includegraphics[width=\textwidth]{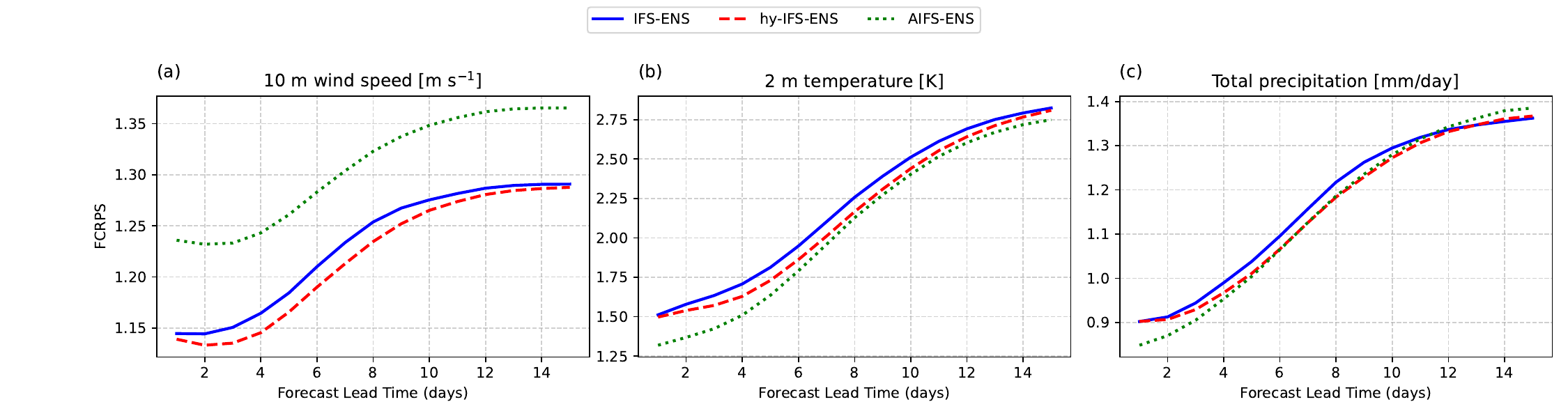}
\caption{Fair CRPS of (a) 10-m wind speed, (b) 2-m temperature and (c) total precipitation for hy-IFS-ENS (dashed red), IFS-ENS (solid blue) and operational AIFS-ENS (dotted green) over the Northern Hemisphere. For differences between IFS-ENS and AIFS-ENS, the data here correspond to the lines labelled ``10m wind speed", ``2 meter temperature" and ``Total precipitation" in Figure~\ref{fig:fig2}. Scores are computed against SYNOP surface observations from 207 forecasts initialized at 00 UTC between 20 October 2025 and 12 May 2026.}
\label{fig:fig4}
\end{figure}

To examine the large-scale skill gains in more detail and to place the hy-IFS-ENS results in the context of the pure machine-learned AIFS-ENS, Figures~\ref{fig:fig3} and \ref{fig:fig4} present selected upper-air and surface verification for the three systems. Overall, upper-air scores from operational AIFS-ENS and hy-IFS-ENS are very similar throughout the troposphere, with both clearly outperforming IFS-ENS. The skill gains of both hy-IFS-ENS and the pure AIFS-ENS relative to IFS-ENS are largest in the tropics, where predictive skill is extended by up to about two days. The hy-IFS-ENS is somewhat less skilful than the pure AIFS-ENS for tropical temperature, for reasons that remain to be understood. 

For near-surface variables over the Northern Hemisphere (Figure~\ref{fig:fig4}), hy-IFS-ENS shows improved probabilistic skill relative to IFS-ENS for 10-m wind speed across all lead times. Both the physics-based and hybrid ensembles substantially outperform AIFS-ENS for 10-m wind verification. The weaker performance of AIFS-ENS for near-surface wind is likely linked to its lower horizontal resolution (28 km versus 9 km), consistent with \cite{LangAIFSENS2024}, who showed that 10-m wind verification for AIFS-ENS is considerably better on the N320 grid than on the O96 grid (their Figures~4 and 5). For 2-m temperature, hy-IFS-ENS is slightly better than IFS-ENS, while AIFS-ENS exhibits substantially higher skill than both, especially in the first week of the forecast. This behaviour is consistent with the fact that AIFS-ENS is trained on ERA5 and ECMWF operational land-analysis 2-m temperature fields, and therefore learns to forecast 2-m temperature directly. As a result, achieving skill comparable to AIFS-ENS for 2-m temperature is challenging for hy-IFS-ENS, in which only large-scale upper-air fields are nudged and land–surface processes are not directly constrained. For total precipitation, hy-IFS-ENS shows consistent improvements relative to IFS-ENS of up to about 3\%. Similar overall behaviour is found in other regions (see Figure~\ref{fig:fig2} for hy-IFS-ENS vs. IFS-ENS). For 10-m wind, conclusions are consistent across domains. For 2-m temperature, the advantage of AIFS-ENS is larger in the tropics and smaller over Europe and at longer lead times. For total precipitation, hy-IFS-ENS consistently improves over IFS-ENS across regions, while relative performance against AIFS-ENS varies: AIFS-ENS performs better in the tropics and Southern Hemisphere, whereas hy-IFS-ENS becomes more skilful at longer lead times over Europe.

\begin{figure}
\centering
\includegraphics[width=\textwidth]{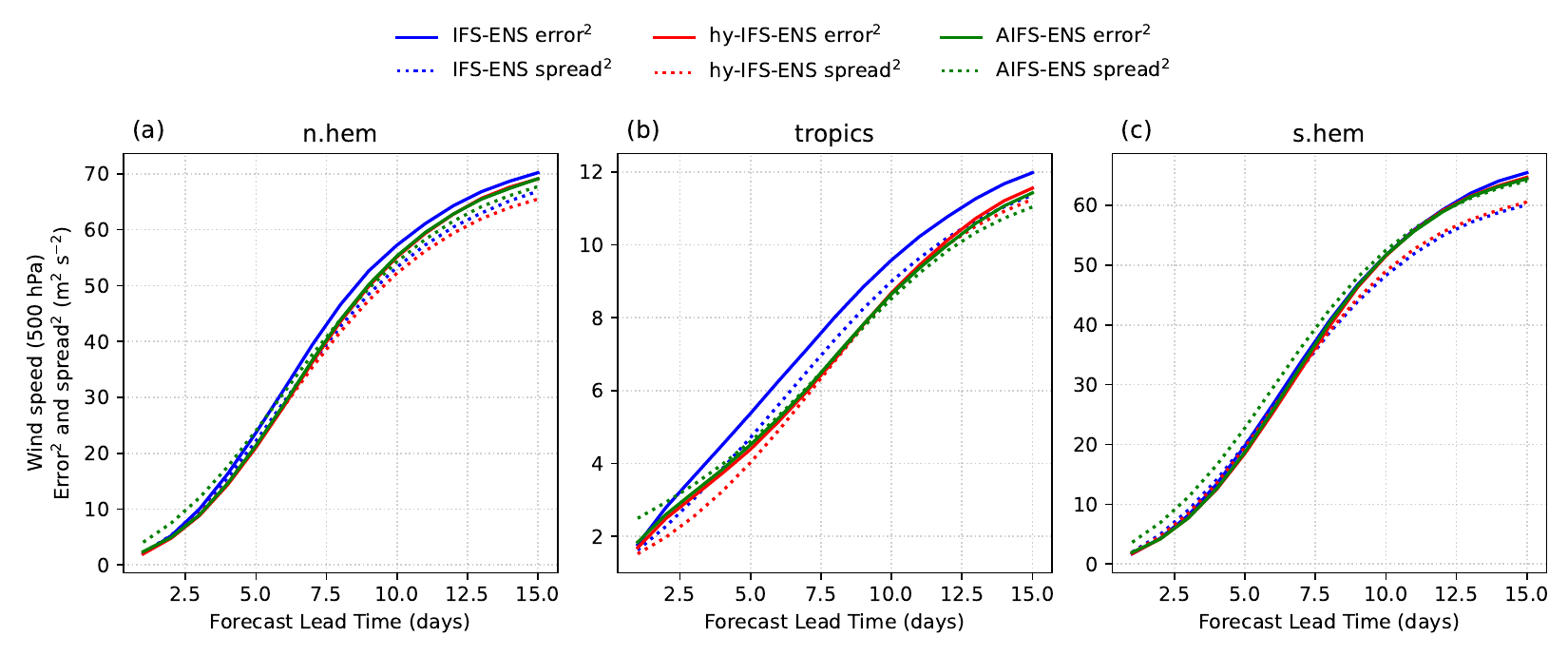}
\caption{Ensemble-mean squared error (solid lines) and ensemble variance (squared ensemble spread; dashed lines) of 500~hPa wind speed for IFS-ENS (blue), hy-IFS-ENS (red), and AIFS-ENS (green) as a function of forecast lead time over the (a) Northern Hemisphere, (b) tropics, and (c) Southern Hemisphere. Spread is scaled by $(N+1)/(N-1)$, where $N$ is ensemble size, to account for finite ensemble size. Error with respect to the ECMWF analysis and ensemble spread are computed from 207 forecasts initialized at 00 UTC between 20 October 2025 and 12 May 2026.}
\label{fig:fig5new}
\end{figure}

To further assess whether the improvement in probabilistic skill in hy-IFS-ENS arises from reduced forecast error rather than changes in ensemble spread, Figure~\ref{fig:fig5new} shows the ensemble-mean squared error and the ensemble variance (spread squared) for 500~hPa wind speed as a function of forecast lead time. For both IFS-ENS and hy-IFS-ENS, ensemble variance grows with lead time in a manner broadly consistent with error growth, although both configurations are under-dispersive in the tropics at short lead times. The hy-IFS-ENS exhibits systematically lower forecast error in all regions for this variable. Ensemble variance is reduced by a comparable but smaller amount, so that the spread-to-error ratio is largely preserved and, from about day~six onwards, brought slightly closer to unity. The improvement in probabilistic skill therefore results primarily from reduced large-scale forecast error, with a
smaller secondary contribution from improved ensemble dispersion at medium and long lead times. 

AIFS-ENS achieves reductions in ensemble-mean squared error relative to IFS-ENS that are comparable to those of hy-IFS-ENS, but its ensemble dispersion behaves differently: it is substantially over-dispersive at short lead times in all three regions. This excess spread decays with lead time, so that beyond about day~10 AIFS-ENS has the spread-to-error ratio closest to unity of the three configurations. Note that hy-IFS-ENS is nudged towards AIFS-ENS-ML, and not towards the AIFS-ENS shown here. The nudging also constrains only wavenumbers below 21, while the verification retains more resolved scales; AIFS-ENS has larger ensemble spread than IFS-ENS and hy-IFS-ENS at total wavenumbers larger than 21 (not shown), its larger spread at short lead times could be dominated by these smaller scales.

\begin{figure}
\centering
\includegraphics[width=\textwidth]{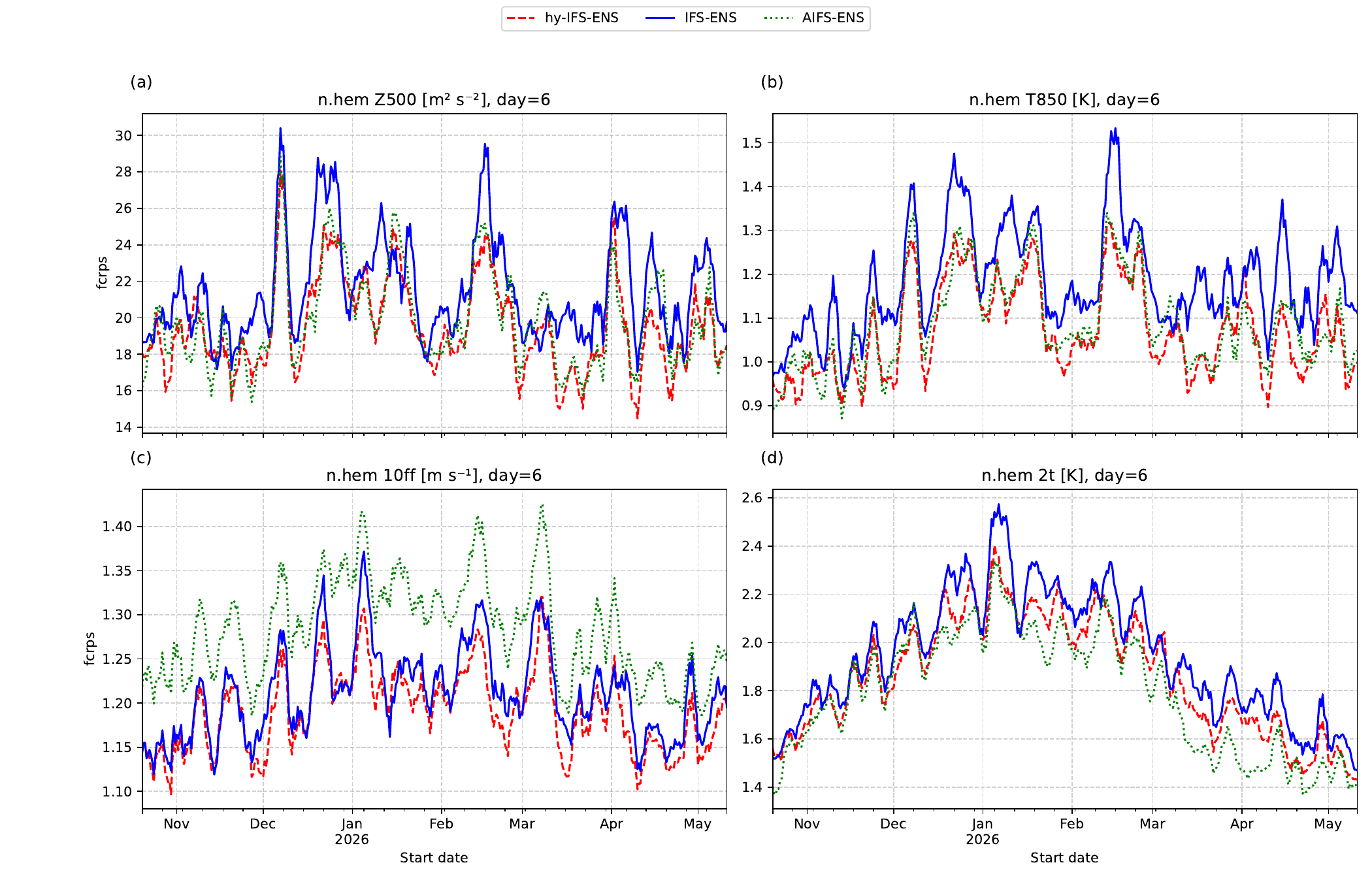}
\caption{Time series of fair continuous ranked probability score (FCRPS) at forecast day 6 for (a) 500~hPa geopotential height (against ECMWF analysis), (b) 850~hPa temperature (against ECMWF analysis), and (c) 10-m wind speed (against SYNOP observations), and (d) 2-m temperature (against SYNOP observations) in the Northern Hemisphere. Scores are shown for IFS-ENS (dashed red), hy-IFS-ENS (solid blue), and operational AIFS-ENS (dotted green). }
\label{fig:fig5}
\end{figure}

Finally, Figure~\ref{fig:fig5} shows the temporal evolution of the fair CRPS at forecast day 6 for Z500, T850 (both verified against ECMWF analysis), and 10-m wind speed, and 2-m temperature (both verified against SYNOP station observations) over the Northern Hemisphere. For all variables considered, hy-IFS-ENS maintains consistently lower (that is, better) FCRPS values than IFS-ENS throughout the evaluation period, indicating that the skill gains associated with spectral nudging are persistent in time rather than confined to specific flow regimes or isolated episodes. The relative ordering of the three systems remains stable. For upper-air fields, hy-IFS-ENS closely tracks AIFS-ENS while retaining a clear advantage over IFS-ENS (Figure~\ref{fig:fig5}a,b). For near-surface fields, hy-IFS-ENS consistently outperforms IFS-ENS and AIFS-ENS for 10-m wind speed, whereas for 2-m temperature the hy-IFS-ENS improves relative to IFS-ENS, while AIFS-ENS maintains a substantial lead over both. This temporal consistency shows that the improvements seen in the summary scorecard at various tropospheric levels are robust and reflect a systematic enhancement of forecast skill rather than being isolated to particular events or periods.

\subsection{Tropical cyclones}\label{section:TCs}
In cases where deterministic physics-based forecasts are spectrally nudged on large scales to machine-learned forecasts, tropical cyclone tracks improve without degrading tropical cyclone intensity forecasts \citep{husain2024leveraging,Polichtchouk2025,niu2025machine}. This improvement is attributed to enhanced representation of the large-scale steering flow in the nudged model. It is therefore insightful to examine whether similar benefits extend to the ensemble configuration.

\begin{figure}
\centering
\includegraphics[width=\textwidth]{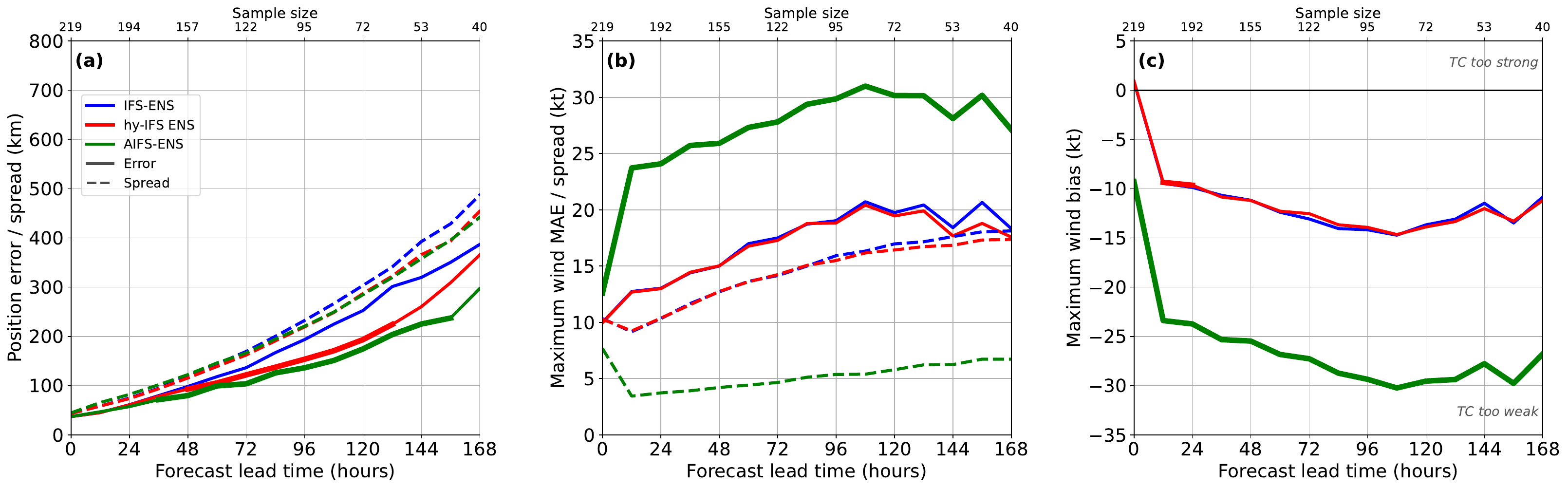}
\caption{Mean position error (a), mean absolute maximum wind error (b), and mean maximum wind bias (c) for tropical cyclones as a function of forecast lead time for the hy-IFS-ENS, IFS-ENS, and AIFS-ENS. Solid lines show the ensemble-mean error, dashed lines the ensemble spread. Verification statistics are computed with respect to IBTrACS for forecasts initialised at 00 UTC between 20 October 2025 and 12 May 2026. Case counts for each lead time are displayed above the graphs. Only common cases across all three configurations are considered. Negative values in (c) indicate that the ensemble-mean maximum wind is too weak with respect to the best track. Thicker line segments indicate lead times where the difference with respect to IFS-ENS is statistically significant at the 5\% level (two-sided paired t-test).}
\label{fig:fig6}
\end{figure}

The hy-IFS-ENS exhibits the same beneficial behaviour. As shown in Figure~\ref{fig:fig6}b-c, tropical cyclone intensity is well preserved: the hy-IFS-ENS closely follows the maximum wind speed evolution of the IFS-ENS, when verified against International Best Track Archive for Climate Stewardship (IBTrACS) data \citep{Gahtan,knapp2010international}, with comparable ensemble-mean errors and ensemble spread. This indicates that spectral nudging does not degrade the intensity characteristics of the IFS. Both systems underestimate observed intensities by roughly 10-15~kt beyond 12h (Figure~\ref{fig:fig6}c), which is a known limitation of 9~km resolution. Increasing resolution to 4.4~km would be expected to improve intensity forecasts \citep[e.g.,][]{majumdar2023advanced,PolichtchoukSanabia2025}. In contrast, the AIFS-ENS under-predicts tropical cyclone intensity far more strongly, by 23-30~kt beyond 12~h, consistent with its lower horizontal resolution. Its ensemble spread in maximum wind is moreover much smaller than its error, whereas spread and error are more comparable in the IFS-ENS and hy-IFS-ENS.

At the same time, tropical cyclone track forecasts are clearly improved in the hy-IFS-ENS. Figure~\ref{fig:fig6}a shows that mean position errors are consistently smaller for hy-IFS-ENS than for IFS-ENS, with the largest differences between forecast day 2-6. This corresponds to a gain of roughly 12--24h of track predictability between forecast days 3 and 6. Importantly, this improvement is achieved without a systematic reduction in ensemble spread, indicating that the improvement arises primarily from reduced large-scale forecast error rather than a collapse of ensemble variability at the mesoscale. This behaviour nevertheless is consistent with a tendency toward under-confidence in hy-IFS-ENS, suggesting that future developments should aim to improve spread–error consistency while retaining the track forecast benefit. Overall, these results indicate that spectral nudging toward AIFS-ENS-ML improves the large-scale circulation in a way that enhances tropical cyclone track skill while maintaining realistic storm intensity evolution, consistent with the large-scale skill improvements discussed in Section~\ref{sec:large-scale}. Finally, the AIFS-ENS has the smallest position errors of the three systems, significantly smaller than those of the IFS-ENS from forecast day 1.5 to day 6.5 and corresponding to a gain of roughly 12--36h of track predictability, although its storm intensities are underestimated.

\subsection{Northern Hemisphere Extremes}\label{sec:NHextremes}
In addition to tropical cyclones, we assess the impact of hybrid forecasting on near-surface extremes over the Northern Hemisphere extra-tropics using SYNOP surface observations. Extremes are quantified through an analysis of the distribution tails of near-surface variables, using probability density functions for 10-m wind speed and 2-m temperature, and exceedance probabilities for 24-h accumulated precipitation. It should be noted that these statistics are also influenced by representativeness effects. SYNOP observations are point measurements, whereas model values represent grid-box averages. In addition, IFS-ENS and hy-IFS-ENS are run at higher horizontal resolution than AIFS-ENS, which can further affect the tails of the distribution.

\begin{figure}
 \centering
 \includegraphics[width=\textwidth]{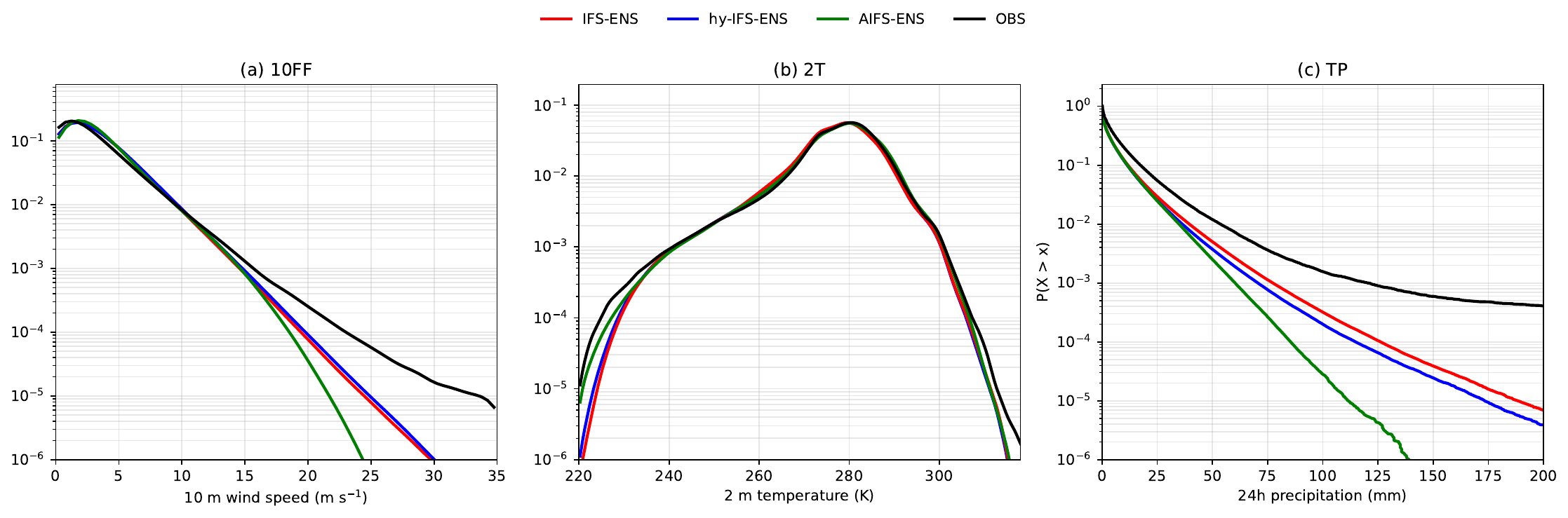}
 \caption{Probability distributions of (a) 10-m wind speed, (b) 2-m temperature, and (c) 24-h accumulated precipitation at forecast day 5 over the Northern Hemisphere for IFS-ENS, hy-IFS-ENS, AIFS-ENS, and SYNOP observations, respectively. Wind speed and temperature are shown as probability density functions obtained by histogramming with 0.5-unit bins (m/s
  and K, respectively) and smoothing with a Gaussian convolution kernel ($\sigma=1.0$ units).  Precipitation is presented as exceedance probability $P(X > x)$. All distributions are computed from station-matched samples at common verification times. 
}
 \label{fig:fig8}
 \end{figure}

Figure~\ref{fig:fig8} shows differences between the experiments primarily in the tails of the distributions, corresponding to extreme events, shown here at a forecast lead time of 5 days. Similar behaviour is found at other lead times (not shown). For 10-m wind speed (Figure~\ref{fig:fig8}a), observations exhibit a heavier tail than all forecast systems, indicating that strong wind events occur more frequently in reality than in the forecasts. IFS-ENS and hy-IFS-ENS show very similar behaviour, both underestimating the frequency of high wind speeds, while AIFS-ENS exhibits the strongest underestimation, with a more rapid decay beyond approximately 15–20~m~s$^{-1}$. This behaviour suggests a general under-representation of intense surface winds in the ensemble forecasts.

For 2-m temperature (Figure~\ref{fig:fig8}b), differences are primarily confined to the cold extremes. All systems underestimate cold extremes, but differences between IFS-ENS and hy-IFS-ENS remain small, indicating little impact of hybrid forecasting on temperature extremes at this lead time. In contrast, AIFS-ENS represents cold extremes most accurately, consistent with its improved large-scale 2-m temperature skill (Figure~\ref{fig:fig4}b). The underestimation of near-surface cold extremes in global physics-based NWP models is often attributed to difficulties in representing stable boundary layers and surface inversions \citep[e.g.,][]{sandu2013so}.

The precipitation exceedance distributions (Figure~\ref{fig:fig8}c) show the largest differences in the upper tail. Heavy precipitation events occur more frequently in observations than in all forecast systems, indicating a general underestimation of extreme accumulations. IFS-ENS and hy-IFS-ENS behave very similarly, while AIFS-ENS shows the strongest deficit of extreme precipitation, likely due to its lower horizontal resolution. Overall, the results suggest that hybrid forecasting preserves the statistics of near-surface extremes relative to IFS-ENS, whereas AIFS-ENS tends to underestimate wind and precipitation extremes while showing comparatively better performance for 2-m temperature.

\section{Summary, Discussion and Conclusions}\label{sec:conclusion}
This study presented the first application of hybrid forecasting using spectral nudging in a probabilistic ensemble forecasting framework. Spectral nudging has previously been demonstrated in deterministic hybrid configurations \citep{husain2024leveraging,Polichtchouk2025,niu2025machine}, but its extension to ensemble prediction requires additional care, as constraining the large-scale flow directly interacts with ensemble spread generation through initial perturbations and stochastic physics schemes such as SPP. If applied naively, tests showed that this can lead to under-dispersion by suppressing physically meaningful large-scale variability. 

The main findings of this study are as follows: 
First, spectral nudging of the physics-based IFS ensemble toward the machine-learned AIFS-ENS ensemble leads to substantial improvements in large-scale forecast skill relative to the un-nudged IFS ensemble. The large-scale skill gains of the pure machine-learned ensemble are effectively transferred to the hybrid system, particularly for upper-air tropospheric variables. Here, the largest relative improvements occur in the tropics, where forecast skill is improved by up to approximately two days, while improvements of around half a day of forecast skill are obtained in both the Northern and Southern Hemisphere extra-tropics.

Second, although nudging is applied only to upper-air large-scale fields, improvements are also seen in several near-surface parameters, such as 2-m temperature, 10-m wind speed and precipitation. This reflects the improved large-scale flow and its influence on surface weather, even though land–surface and boundary-layer processes are not directly constrained by the nudging.

Third, the hybrid ensemble shows a clear improvement in tropical cyclone track forecasts, with mean position errors reduced by up to approximately 1 day, while tropical cyclone intensity forecasts remain unchanged. This shows that improvements in large-scale steering flow can be achieved without degrading the mesoscale structure of the storms, which continues to benefit from the higher spatial resolution of the physics-based IFS ensemble.

Hybrid ensemble forecasting combines machine-learned large-scale skill with the spatio-temporal resolution and physically based dynamics of the IFS. It can be interpreted as a scale-selective, flow-dependent reduction of large-scale forecast error: spectral nudging transfers predictive skill from the machine-learned ensemble to the physics-based model by relaxing its large-scale components toward a higher-skill reference forecast, while scales below the nudging cut-off continue to evolve according to the model dynamics and physical parametrizations. No instability or episodic forecast failures are observed, and the hybrid ensemble maintains robust skill across a wide range of variables. Compared to AIFS-ENS, the hybrid system retains the full set of IFS output parameters and a comparable range of forecast products.

The hybrid framework also provides a pathway for generating physically based high-resolution forecasts constrained by machine-learned information. In this sense, the hybrid integrations will also be useful for training future machine-learned down-scaling approaches, potentially enabling high-resolution ensemble products with reduced reliance on the full physics-based model in the time-critical forecasting path.

The application of spectral nudging increases the computational cost of the IFS ensemble integration by approximately 9\%. This increase is in addition to the cost associated with running the AIFS-ENS-ML inference and preparing the fields used as the nudging reference. However, the AIFS-ENS-ML configuration used here is computationally cheap and 15 forecast days take less than one minute per ensemble member on an NVIDIA A100 GPU.

The additional expense for running the nudged IFS arises primarily from applying the nudging below the diagnosed tropopause, which must be determined in grid-point space at every model time step. As a result, additional spectral-to-grid-point transforms are required in order to compute and apply the nudging tendencies. An alternative approach would be to apply nudging only below a pre-defined model level, which avoids the need for additional spectral-to-grid-point transforms and reduces the cost increase to approximately 1.5\%. Such an approach may be attractive if computational cost is a primary constraint. However, because the tropopause height varies substantially with latitude—occurring near 100~hPa in the tropics and closer to 300~hPa in polar regions, a fixed-level configuration would require nudging to be limited to levels below approximately 300~hPa in order to avoid degrading the stratospheric circulation. This would remove a substantial part of the nudging influence in the tropical upper troposphere, where some of the largest skill improvements are obtained. Additional cost reductions are also possible within the tropopause-based configuration by applying time-varying nudging, in which the relaxation time-scale varies smoothly in time following a cosine-bell profile as proposed by \cite{husain2024leveraging,husain2014extended}. In this approach, nudging is concentrated around the 6-hourly inference times of AIFS-ENS, while intermediate model time steps are only weakly nudged or left un-nudged, reducing the number of required transforms. The configuration adopted here therefore represents a trade-off between computational efficiency and forecast skill, with continuous tropopause-based nudging retaining the full tropical upper-tropospheric benefit while remaining computationally feasible.

From a research-to-operations perspective, the transition to hybrid forecasting is attractive, since spectral nudging can be integrated into existing NWP workflows with potentially little impact on the development of the physics-based model. However, some practical challenges remain. For example, maintaining consistency between real-time forecasts and hindcasts used for calibration and products such as the Extreme Forecast Index (EFI) requires consideration, because the hindcast period has been used for training the machine-learned ensemble system. However, tests in which the hybrid EFI is calculated using the un-nudged IFS hindcast climatology are encouraging, with the hybrid EFI showing improved skill relative to the IFS-ENS EFI despite the use of the un-nudged climatology (not shown). In addition, there might be a need to retrain the machine-learned component when the physics-based forecasting system, and in particular the analysis, changes. At the same time, the hybrid framework retains flexibility, since the nudging can be adjusted or disabled if required, allowing a direct fallback to the baseline physics-based ensemble configuration. The coexistence of pure machine-learned, physics-based, and hybrid configurations may also make future comparisons between machine-learned and physics-based forecasting systems less straightforward, as hybrid systems blur the distinction between the two paradigms.

While the present results are obtained at 9 km resolution, the hybrid configuration is not resolution-specific and not limited by the resolution of the machine-learned forecasting model. Therefore, we anticipate that similar improvements will extend to kilometre-scale configurations.

\section*{Acknowledgements}  We acknowledge the EuroHPC Joint Undertaking for awarding this work access to the EuroHPC supercomputer MN5, hosted by BSC in Barcelona through a EuroHPC JU Special Access call. We thank Michail Diamantakis, Stephen English and Martin Leutbecher for providing useful comments on the manuscript and Christopher Roberts for useful discussions. 

\section*{Conflict of interest statement}
The authors declare no conflict of interest.

\section*{Data availability statement}
The hy-IFS-ENS forecast experiment used in this study is publicly available at: https://apps.ecmwf.int/ifs-experiments/rd/iy2u/. The operational IFS analysis and IFS-ENS data are available at https://www.ecmwf.int/en/forecasts/datasets/open-data under ECMWF’s open data
policy.

Model IFS code developed at ECMWF are the intellectual
property of ECMWF and its member states, and therefore is not publicly available. Access to a reduced version of the IFS code may be obtained from ECMWF
under an OpenIFS licence (see http://www.ecmwf.int/en
/research/projects/openifs for further information).

The code for AIFS-ENS is available at https://github.com/ecmwf/
anemoi-core.

\bibliographystyle{wileyqj}
\bibliography{References}

\end{document}